\documentclass[a4paper]{article}
\usepackage{a4wide}
\usepackage[UKenglish]{babel}
\usepackage{amsmath,amssymb,amsthm,mathtools,empheq,mathrsfs,bbold}
\usepackage{bbm}
\usepackage{graphicx,subfigure}
    \graphicspath{{./}{figs/}}
\usepackage{xcolor}
\usepackage[colorlinks=true,linkcolor=blue,citecolor=red]{hyperref}
\usepackage{authblk}
\usepackage[inline]{enumitem}

\DeclarePairedDelimiter\abs{\lvert}{\rvert}

\newcommand{\bbA}{\mathbf{A}}
\newcommand{\bbc}{\mathbf{c}}
\newcommand{\cC}{\mathcal{C}}
\newcommand{\bbf}{\mathbf{f}}

\newcommand{\bbg}{\mathbf{g}}
\newcommand{\cG}{\mathcal{G}}
\newcommand{\N}{\mathbb{N}}
\newcommand{\norm}[2]{\Vert#1\Vert_{#2}}
\newcommand{\pr}[1]{{}^\prime\!#1}
\newcommand{\R}{\mathbb{R}}
\newcommand{\brho}{\boldsymbol{\rho}}
\newcommand{\cW}{\mathcal{W}}

\newtheorem{proposition}{Proposition}[section]
\newtheorem{theorem}[proposition]{Theorem}

\theoremstyle{remark}

\allowdisplaybreaks

\title{Homogeneous Boltzmann-type equations on graphs: A framework for modelling networked social interactions}

\author{Andrea Tosin}
\affil{{\small Department of Mathematical Sciences ``G. L. Lagrange'' \\ Politecnico di Torino, Italy}}
\date{}

\begin{document}
\maketitle

\begin{abstract}
Homogeneous Boltzmann-type equations are an established tool for modelling interacting multi-agent systems in sociophysics by means of the principles of statistical mechanics and kinetic theory. A customary implicit assumption is that interactions are ``all-to-all'', meaning that every pair of randomly sampled agents may potentially interact. However, this legacy of classical kinetic theory, developed for collisions among gas molecules, may not be equally applicable to social interactions, which are often influenced by preferential connections between agents. In this paper, we discuss ongoing research on incorporating graph structures into homogeneous Boltzmann-type equations, thereby accounting for the ``some-to-some'' nature of social interactions.

\medskip

\begin{center}
\textbf{Sommario}
\end{center}
Le equazioni di tipo Boltzmann omogenee sono uno strumento consolidato per modellizzare sistemi multi-agente nella sociofisica mediante i princ\`{i}pi della meccanica statistica e della teoria cinetica. Un'ipotesi implicita ricorrente \`{e} che le interazioni siano ``tutti con tutti'', cio\`{e} che ogni coppia di agenti selezionati casualmente possa interagire. Tuttavia, questo retaggio della teoria cinetica classica, sviluppata per descrivere gli urti tra molecole di gas, non \`{e} probabilmente altrettanto adeguato per descrivere le interazioni sociali, che spesso risentono di connessioni preferenziali tra gli agenti. In quest'articolo discutiamo alcune linee di ricerca attualmente in corso sull'integrazione di strutture a grafo nelle equazioni di tipo Boltzmann omogenee, con l'obiettivo di tenere conto di interazioni sociali di tipo ``alcuni con alcuni''.

\bigskip


\medskip

\end{abstract}

\section{Introduction}
The \textit{homogeneous Boltzmann equation}, a special case of the integro-differential equation introduced by Ludwig Boltzmann at the end of the 19th century, describes the statistics of elastic collisions among the molecules of a rarefied gas homogeneously distributed in space in accordance to the principles of the \textit{kinetic theory}. Gas molecules are regarded as \textit{indistinguishable}, so that any of them is representative of all the molecules of the gas, and are characterised by their velocity $v\in\R^3$, which changes in consequence of the collisions. Only \textit{binary}, i.e. pairwise, collisions are considered, assuming that events of simultaneous collisions among more than two molecules are negligible.

If $v,\,v_\ast\in\R^3$ are the instantaneous velocities of two colliding molecules, the result of a collision between them is that the \textit{pre-collisional} velocities $v,\,v_\ast$ change into \textit{post-collisional} velocities $v',\,v_\ast'\in\R^3$, which are given by the physical laws of elastic collisions (those based on the conservation of momentum and kinetic energy of the two molecules):
\begin{equation}
	v'=v+[(v_\ast-v)\cdot n]n, \qquad v_\ast'=v_\ast+[(v-v_\ast)\cdot n]n,
	\label{eq:direct_coll}
\end{equation}
where $n\in\mathbb{S}^2$ is a unit normal vector in the direction joining the centres of the colliding molecules, also termed the direction of collision, and $\cdot$ denotes the Euclidean inner product in $\R^3$.

The superposition of a large number of collisions of type~\eqref{eq:direct_coll} in the unit time causes a change in time of the statistical distribution of the velocities of the molecules. Notice that the reference to a statistical description is essential, for it would be otherwise impossible to track deterministically the collisions among all pairs of gas molecules at every time. For this, one introduces the \textit{distribution function} $f=f(v,t):\R^3\times (0,\,+\infty)\to\R_+$, such that $f(v,t)\,dv$ yields the probability that at time $t$ a representative molecule of the gas has a velocity in the infinitesimal volume $dv$ of the state space $\R^3$ centred at $v$. In other words, for every measurable (e.g., Borel) set $A\subseteq\R^3$ it results
$$ \operatorname{Prob}(V_t\in A)=\int_A f(v,t)\,dv, $$
$V_t$ being the random variable expressing the velocity of a representative gas molecule at time $t$. Moreover, the normalisation condition $\int_{\R^3}f(v,t)\,dv=1$ holds for every $t\geq 0$.

The homogeneous Boltzmann equation is an integro-differential equation expressing the time evolution of $f$ under the collision rules~\eqref{eq:direct_coll}. The equation writes as
\begin{equation}
	\frac{\partial f}{\partial t}(v,t)=\frac{1}{4\pi}\int_{\R^3}\int_{\mathbb{S}^2}B((v_\ast-v)\cdot n)\bigl(f(\pr{v},t)f(\pr{v}_\ast,t)-f(v,t)f(v_\ast,t)\bigr)\,dn\,dv_\ast
	\label{eq:homog_Boltz}
\end{equation}
and, in practice, equates the time variation of $f$ (on the left-hand side) to the mean effect of the elastic collisions~\eqref{eq:direct_coll} in the unit time (on the right-hand side). For a given velocity $v$, such a mean effect is evaluated by computing statistically the net balance between:
\begin{enumerate}[label=(\roman*)]
\item The collisions leading a molecule to gain $v$ as post-collisional velocity starting from whatever pre-collisional velocity $\pr{v}$, thereby determining an increase in time of $f$. This is expressed by the first term on the right-hand side of~\eqref{eq:homog_Boltz}, called the \textit{gain term}:
$$ \frac{1}{4\pi}\int_{\R^3}\int_{\mathbb{S}^2}B((v_\ast-v)\cdot n)f(\pr{v},t)f(\pr{v}_\ast,t)\,dn\,dv_\ast, $$
which is an average over all possible directions of collision $n$ (notice that $\frac{1}{4\pi}$ is the Hausdorff measure of $\mathbb{S}^2$, motivated by the implicit assumption that $n$ is uniformly distributed in $\mathbb{S}^2$) and all possible post-collisional velocities $v_\ast$ of the other molecule participating in the collision.

Here, $\pr{v},\,\pr{v}_\ast\in\R^3$ have to be understood as functions of the post-collisional velocities $v,\,v_\ast$. More precisely, they are the pre-collisional velocities which, after a collision~\eqref{eq:direct_coll}, generate the post-collisional velocities $v,\,v_\ast$. As such, they are obtained by inverting the collision rules~\eqref{eq:direct_coll}, which results in
\begin{equation}
	\pr{v}=v+[(v_\ast-v)\cdot n]n, \qquad \pr{v}_\ast=v_\ast+[(v-v_\ast)\cdot n]n.
	\label{eq:inv_coll}
\end{equation}
These new rules are called the \textit{inverse collisions}. Notice that they are exactly the same as~\eqref{eq:direct_coll} with the roles of the pre-collisional and post-collisional velocities reversed. In particular, it is important to pay attention to the fact that, while in~\eqref{eq:direct_coll} $v,\,v_\ast$ denote the pre-collisional velocities, in~\eqref{eq:inv_coll} they denote the post-collisional velocities.

\item The collisions leading a molecule having $v$ as pre-collisional velocity to lose it, thereby causing a decrease in time of $f$. This is expressed by the second term on the right-hand side of~\eqref{eq:homog_Boltz}, called the \textit{loss term}:
$$ -\frac{1}{4\pi}\int_{\R^3}\int_{\mathbb{S}^2}B((v_\ast-v)\cdot n)f(v,t)f(v_\ast,t)\,dn\,dv_\ast, $$
which is in turn an average over all possible directions of collision $n$ and all possible velocities $v_\ast$ of the other colliding molecule.
\end{enumerate}

The sum of gain and loss terms gives the so-called \textit{collisional operator} $Q$:
$$ Q(f,f)(v,t):=\frac{1}{4\pi}\int_{\R^3}\int_{\mathbb{S}^2}B((v_\ast-v)\cdot n)\bigl(f(\pr{v},t)f(\pr{v}_\ast,t)-f(v,t)f(v_\ast,t)\bigr)\,dn\,dv_\ast, $$
a bilinear integral operator conferring on the homogeneous Boltzmann equation its typical \textit{integro}-differential character.

The term $B$ is the \textit{collision kernel}. It accounts for the rate of collision between pairs of molecules with given pre-collisional velocities and, in particular, relates such a rate to the alignment of their relative velocity $v_\ast-v$ to the direction of collision $n$. Common expressions of $B$ are even functions, such as
$$ B(\nu)=\abs{\nu} \quad \text{whence} \quad B((v_\ast-v)\cdot n)=\abs{(v_\ast-v)\cdot n}, $$
which implies that the higher the said alignment the higher the rate of collision. Another widely used model is $B\equiv 1$, i.e. a constant rate of collision, independent of the relative velocity, which characterises the so-called \textit{Maxwellian molecules}.

For a thorough derivation of the homogeneous Boltzmann equation from molecule collisions, we refer the interested reader to~\cite{Loy2025}.

Starting from the early $2000$s, the homogeneous Boltzmann equation has been taken systematically as a paradigm to model systems of interacting particles quite different from gas molecules. Very often, such ``particles'' are not even particles in the sense of classical physics. Instead, they may be vehicles along a road in car traffic problems~\cite{Prigogine1960,Tosin2019}, human beings in opinion formation problems~\cite{Boudin2009,Toscani2006} or trading entities in wealth distribution problems~\cite{Bisi2017,Cordier2005}, to mention just a few examples. In such cases, they are preferentially called \textit{agents} and one speaks of \textit{interacting multi-agent systems}~\cite{Pareschi2013}.

The characteristic trait, still denoted by $v$, of a representative agent of one of such systems is not necessarily the velocity like for gas molecules. Moreover, it is typically assumed to be scalar instead of vector-valued, hence $v\in\R$. In car traffic problems, $v$ is the speed of the vehicles; in opinion formation problems, it is the opinion of the individuals about a certain topic, with $v=0$ representing neutrality and $v>0$, $v<0$ opposite convictions; in wealth distribution problems, it is the wealth of the individuals, with $v<0$ possibly standing for debts~\cite{Torregrossa2018}.

An interaction between two representative agents of the system with pre-interaction traits $v,\,v_\ast$ consists in a transformation of $v,\,v_\ast$ into post-interaction traits $v',\,v_\ast'$ according to prescribed relationships -- the \textit{interaction rules} -- expressing $v',\,v_\ast'$ as functions of $v,\,v_\ast$. In this context, the interaction rules are the counterpart of the collision rules~\eqref{eq:direct_coll} in modelling the ``physics'' of the multi-agent system.

Motivated by the linearity and symmetry of the collision rules~\eqref{eq:direct_coll} -- the symmetry meaning that either rule can be deduced from the other by swapping the roles of $v$ and $v_\ast$ --, here we focus on general scalar linear symmetric interaction rules of the form
\begin{equation}
	v'=pv+qv_\ast, \qquad v_\ast'=pv_\ast+qv
	\label{eq:int_rules}
\end{equation}
with prescribed interaction coefficients $p,\,q\geq 0$. These rules are invertible provided $q\neq p$; under this assumption, the \textit{inverse interactions} are expressed by
\begin{equation}
	\pr{v}=\frac{p}{p^2-q^2}v-\frac{q}{p^2-q^2}v_\ast, \qquad \pr{v}_\ast=\frac{p}{p^2-q^2}v_\ast-\frac{q}{p^2-q^2}v,
	\label{eq:inv_int}
\end{equation}
where $\pr{v},\,\pr{v}_\ast$ are the pre-interaction traits and $v,\,v_\ast$ the post-interaction traits.

The distribution function $f=f(v,t):\R\times (0,\,+\infty)\to\R_+$ of the trait $v$ is introduced in such a way that $f(v,t)\,dv$ is the probability that at time $t$ a representative agent has a trait in the infinitesimal interval $dv$ centred at $v$. Such an interval is sometimes identified explicitly with $[v-\frac{dv}{2},\,v+\frac{dv}{2}]$. A homogeneous \textit{Boltzmann-type} equation for the evolution of $f$ can be written as, cf.~\cite{Loy2025,Toscani2006},
\begin{equation}
	\frac{\partial f}{\partial t}(v,t)=\int_\R\left(\frac{B(\pr{v},\pr{v}_\ast)}{\abs{p^2-q^2}}f(\pr{v},t)f(\pr{v}_\ast,t)-B(v,v_\ast)f(v,t)f(v_\ast,t)\right)dv_\ast,
	\label{eq:homog_Boltz-type}
\end{equation}
where $\pr{v},\,\pr{v}_\ast$ are understood as functions of $v,\,v_\ast$ through~\eqref{eq:inv_int} and $\frac{1}{\abs{p^2-q^2}}$ is the Jacobian factor of the transformation~\eqref{eq:inv_int}. The similarity of~\eqref{eq:homog_Boltz-type} with the homogeneous Boltzmann equation~\eqref{eq:homog_Boltz} is evident, yet two comments are in order:
\begin{enumerate}[label=(\roman*)]
\item in the homogeneous Boltzmann equation, there is apparently no Jacobian factor because that of the transformation~\eqref{eq:inv_coll} is unitary;
\item in the homogeneous Boltzmann equation, an even collision kernel $B$ together with the collision rules~\eqref{eq:direct_coll},~\eqref{eq:inv_coll} implies $B((\pr{v}_\ast-\pr{v})\cdot n)=B((v_\ast-v)\cdot n)$, cf.~\cite{Loy2025}, hence that term can be factored out in the collisional operator.
\end{enumerate}

It is worth mentioning that often~\eqref{eq:homog_Boltz-type} is rewritten in weak form to eliminate at once both the Jacobian factor and the inverse interaction. For this, one multiplies both sides of~\eqref{eq:homog_Boltz-type} by an \textit{observable} $\varphi=\varphi(v):\R\to\R$, which plays the role of a test function, and integrates on $\R$ with respect to $v$ to get
\begin{equation}
	\frac{d}{dt}\int_\R\varphi(v)f(v,t)\,dv=\int_\R\int_\R B(v,v_\ast)\bigl(\varphi(v')-\varphi(v)\bigr)f(v,t)f(v_\ast,t)\,dv\,dv_\ast,
	\label{eq:homog_Boltz-type.weak}
\end{equation}
where now $v'$ is given as a function of $v,\,v_\ast$ by the interaction rules~\eqref{eq:int_rules}. Equation~\eqref{eq:homog_Boltz-type.weak} states that the time variation of the average of the observable (left-hand side) is determined by the mean variation of the observable in a representative interaction (right-hand side). Notice indeed that the difference $\varphi(v')-\varphi(v)$ quantifies the variation of $\varphi$ after an interaction, being $\varphi(v),\,\varphi(v')$ the pre-interaction and post-interaction values of the observable, respectively.

The interaction dynamics considered so far implicitly assume that any pair of agents randomly sampled from the entire pool can interact; that is, the agents are homogeneously mixed and thus experience \textit{all-to-all} interactions. This is appropriate for freely moving gas molecules, but may not hold in sociophysical applications, where agents exhibit \textit{preferential connections}. Examples include social or Internet networks, whose agents -- human beings or electronic devices -- interact and exchange information only if they follow/are connected to each other.

In this paper, we present various strategies for including a \textit{network structure} in kinetic descriptions of multi-agent interactions. The basic idea is to introduce a \textit{graph}, which models the connections among the agents, and to embed the information encoded in its \textit{adjacency matrix} into a homogeneous Boltzmann-type equation, so as to discriminate which pairs of agents can actually interact. We also discuss the possibility to let the number of graph vertices and edges grow to infinity, in order to recover, in the limit, a statistical description of the connections valid for large dense graphs, free from the detail of the single connections required by the adjacency matrix.

We denote by $\cG_N$ the graph, where $N\in\N$ is the number of vertices. These are identified by an index $i\in\mathcal{V}_N:=\{1,\,\dots,\,N\}$. Instead, the edges of $\cG_N$ are identified with pairs of vertices belonging to a set $\mathcal{E}_N\subseteq\mathcal{V}_N^2$. In more detail, given $i,\,j\in\mathcal{V}_N$, we say that $(i,\,j)\in\mathcal{E}_N$ if there exists an edge between the vertices $i$ and $j$. Thus, $\cG_N=(\mathcal{V}_N,\,\mathcal{E}_N)$. For simplicity, we confine ourselves to \textit{undirected} graphs, i.e. graphs for which vertices are either reciprocally connected or disconnected. Formally, this means that if $(i,\,j)\in\mathcal{E}_N$ then also $(j,\,i)\in\mathcal{E}_N$. As a consequence, the adjacency matrix of $\cG_N$, which we denote by $\bbA_N\in\R^{N\times N}$, is symmetric.

The remainder of the paper is organised as follows. In Section~\ref{sect:networked_multi-agent}, we consider \textit{networked multi-agent systems}, in which each vertex of $\cG_N$ is a group of interacting agents that can migrate from vertex to vertex following the graph connections. In this case, the number $N$ of vertices remains fixed and finite. In Section~\ref{sect:networked_int}, we discuss instead \textit{networked interactions}. In this case, each vertex of $\cG_N$ is an agent, that may or may not interact with other vertices/agents in accordance with the connections encoded in $\mathcal{E}_N$ or, equivalently, $\bbA_N$. Here, the focus is on letting $N\to\infty$, in line with the large number of agents required in kinetic theory, in order to obtain a coherent \textit{statistical} description of the network structure and its impact on agent interactions. To this end, advanced tools from graph theory, such as random graphs and graphons, are fruitfully integrated into kinetic equations, thereby opening up promising directions for further research.

\section{Networked multi-agent systems}
\label{sect:networked_multi-agent}
\begin{figure}[!t]
\centering
\includegraphics[width=.45\textwidth]{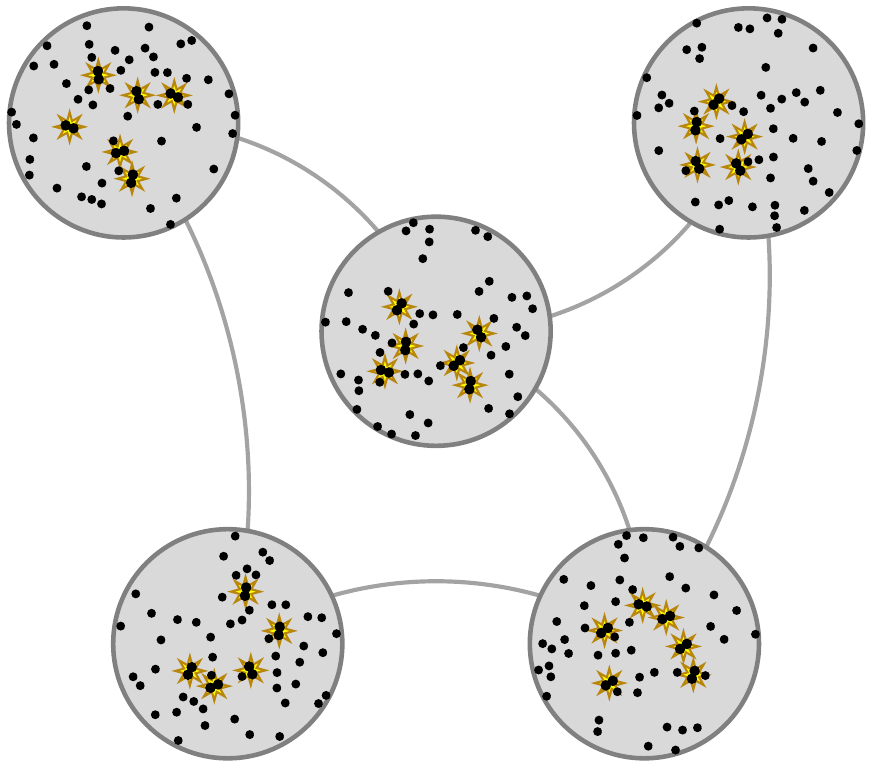}
\caption{Five networked multi-agent systems with binary interactions within each of them}
\label{fig:networked_MAS}
\end{figure}

We begin by considering the case of \textit{networked multi-agent systems}. We understand the latter as a collection of $N$ groups of interacting agents, which may exchange individuals following certain connections among them. The $N$ groups are the vertices of the graph $\cG_N$ mentioned in the introductory section. See Figure~\ref{fig:networked_MAS}.

Such a structure is suited to model situations in which only the agents belonging to the same group can interact, but agents move across the groups bringing their typical traits from group to group. For example, this is the case of the spread of an infectious disease, where the $N$ systems of agents may represent groups of people living in the same city, region, or country, who propagate the disease locally through direct interactions and globally by moving from one place to another~\cite{Loy2021}. In this application, the trait $v$ which characterises each agent in the various groups may represent, for example, the viral load, see e.g.,~\cite{DellaMarca2022,Dellamarca2023,Lorenzi2024}.

We introduce the kinetic distribution function $f_i=f_i(v,t):\R\times (0,\,+\infty)\to\R_+$, which provides the density of individuals in the $i$-th group that at time $t$ feature a trait in the infinitesimal interval $[v-\frac{dv}{2},\,v+\frac{dv}{2}]$ centred at $v$. It is worth stressing that, unlike classical kinetic theory, $f_i$ is \textit{not} a probability density -- nor, in more generality, a probability measure --, because the number of agents in group $i$ is not constant in time, due to the aforesaid migrations of the agents from group to group. This difficulty of the theory is however mitigated by the fact that the quantity
$$ \sum_{i=1}^{N}f_i(v,t) $$
is instead constant in time, because the agents switch from one group to another but they never leave the $N$ groups as a whole. Hence, the total number of agents in the $N$ groups is conserved.

These considerations are at the basis of the procedure which leads to obtain a system of Boltzmann-type equations for the $f_i$'s from the said dynamics of interactions and migrations. The procedure, which is described in detail in~\cite{Loy2021a}, produces
\begin{equation}
	\frac{\partial f_i}{\partial t}=\lambda Q(f_i,f_i)+\chi\left(\sum_{j=1}^{N}a_{ij}f_j-f_i\right), \qquad i=1,\,\dots,\,N.
	\label{eq:Boltz.graph.migr}
\end{equation}

Here, $Q$ is the collisional operator
$$ Q(f_i,f_i)(v,t)=\int_\R\left(\frac{1}{\abs{p^2-q^2}}f_i(\pr{v},t)f_i(\pr{v}_\ast,t)-f_i(v,t)f_i(v_\ast,t)\right)dv_\ast, $$
which expresses the statistics of the interactions taking place, at rate $\lambda>0$, among the agents of group $i$ with the rules~\eqref{eq:inv_int}.

The second term on the right-hand side of~\eqref{eq:Boltz.graph.migr} accounts instead for the migration process which, at rate $\chi>0$, causes the agents to move from one group to another. The $a_{ij}$'s, for $i,\,j=1,\,\dots,\,N$, are the entries of the adjacency matrix $\bbA_N$ of $\cG_N$. In particular, $a_{ij}\in [0,\,1]$ is the probability that, in the unit time, an agent moves from group $j$ to group $i$, so that
\begin{equation}
	\sum_{i=1}^{N}a_{ij}=1, \qquad \forall\,j=1,\,\dots,\,N.
	\label{eq:sum.aij}
\end{equation}
In other words, $\cG_N$ is, in general, a weighted graph with left stochastic adjacency matrix. In the particular case that $a_{ij}\in\{0,\,1\}$, $\cG_N$ is an unweighted graph with $a_{ij}=1$ meaning that vertices $i$ and $j$ are connected, hence that agents can move directly from group $j$ to group $i$, and $a_{ij}=0$ meaning that they are not, hence that agents cannot move directly from group $j$ to group $i$.

If we introduce the mass of agents of group $i$ at time $t$:
$$ \rho_i(t):=\int_\R f_i(v,t)\,dv $$
and we integrate~\eqref{eq:Boltz.graph.migr} with respect to $v$, we discover
\begin{equation}
	\frac{d\rho_i}{dt}=\chi\left(\sum_{j=1}^{N}a_{ij}\rho_j-\rho_i\right), \qquad i=1,\,\dots,\,N,
	\label{eq:rhoi}
\end{equation}
where we have used the fact that $\int_\R Q(f_i,f_i)(v,t)\,dv=0$, because the interactions do not change the number of agents in each group. This shows that, due to the migrations of the agents through the groups, the mass of agents in each group is in general not constant in time, i.e. $\frac{d\rho_i}{dt}\neq 0$. In particular, the variation in time of the mass of agents in a certain group $i$ is determined by a balance between:
\begin{enumerate}[label=(\roman*)]
\item the inflow of agents coming from adjacent groups, expressed by the term $\sum_{j=1}^{N}a_{ij}\rho_j$;
\item the outflow of agents heading to other groups, which is proportional to the mass $\rho_i$ of agents in group $i$.
\end{enumerate}

\subsection{Outline of the basic qualitative theory}
In this section, we outline some basic facts of the qualitative theory of~\eqref{eq:Boltz.graph.migr} as developed in~\cite{Bisoglio2024} and further refined in~\cite{Loy2025}. They concern mainly the large time trend of the solutions to~\eqref{eq:Boltz.graph.migr} and~\eqref{eq:rhoi}, which reveals self-organised configurations of the system emerging, in the long run, from the interactions and the migrations of the agents.

\subsubsection{Mass redistribution on the graph}
\label{sect:mass_redistribution}
The redistribution of the agents in the various groups is governed by~\eqref{eq:rhoi}, which is a self-consistent system of ordinary differential equations in the unknowns $\rho_i$, $i=1,\,\dots,\,N$, decoupled from the original system of kinetic equations~\eqref{eq:Boltz.graph.migr}. This makes it possible to study preliminarily the distribution of the mass of agents on the graph $\cG_N$ as the solution $\brho:=(\rho_1,\,\dots,\,\rho_N)$ to the following system of ODEs:
$$ \frac{d\brho}{dt}=\chi(\bbA_N-\mathbf{I}_N)\brho, $$
which is the vector form of~\eqref{eq:rhoi}, $\mathbf{I}_N$ denoting the $N\times N$ identity matrix.

The system being linear, there exists a unique global solution for every prescribed initial condition $\brho_0:=\brho(0)=(\rho_{0,1},\,\dots,\,\rho_{0,N})\in\R^N$. Such a solution turns out to be physically consistent in the sense stated by the following
\begin{theorem}[See~\cite{Loy2025}] \label{theo:rho}
If $\brho_0\in\R^N_+$, i.e. if the initial masses of agents are non-negative in every vertex of $\cG_N$, then $\rho_i(t)\geq 0$ for all $t>0$ and all $i=1,\,\dots,\,N$.

In particular, if $\rho_{0,i}>0$, i.e. if the initial mass in a given vertex $i$ is non-zero, then $\rho_i(t)>0$ for all $t>0$, i.e. the mass in vertex $i$ is in turn non-zero at every successive time.
\end{theorem}

Summing both sides of~\eqref{eq:rhoi} over $i$ and recalling~\eqref{eq:sum.aij}, we see that
$$ \dfrac{d}{dt}\sum_{i=1}^{N}\rho_i(t)=0. $$
Therefore, it is possible to fix the value of the total mass of agents in $\cG_N$. A frequent reference choice is a unitary initial total mass, which, owing to the property above, implies that the total mass is unitary at all successive times. As a consequence, it results $0\leq\rho_i(t)\leq 1$ for all $t>0$, thus each $\rho_i(t)$ can be understood as the percentage of total mass of agents in vertex $i$ at time $t$.

The big picture of the mass redistribution is completed by its time-asymptotic trend:
\begin{theorem}[See~\cite{Loy2025}] \label{theo:rho.asympt}
Let $\cG_N$ be strongly connected. Then, there exists a unique mass distribution $\brho^\infty:=(\rho^\infty_1,\,\dots,\,\rho^\infty_N)\in\R^N$, with $\rho^\infty_i>0$ for all $i=1,\,\dots,\,N$ and $\sum_{i=1}^{N}\rho^\infty_i=1$, which is a stable and globally attractive equilibrium of~\eqref{eq:rhoi}.
\end{theorem}

The assumption that $\cG_N$ is strongly connected means that every pair of vertices is connected by a path of edges. From the modelling point of view, this implies that every group of agents is reacheable from any other group, i.e. that there are no isolated groups. This assumption turns out to be essential to ensure the validity of a result such as the one of Theorem~\ref{theo:rho.asympt}, which claims the existence of a mass distribution emerging for large times independently of the initial mass distribution. Clearly, the emerging mass distribution $\brho^\infty$ depends instead on the adjacency matrix $\bbA_N$.

\subsubsection{Trait distribution on the graph}
System~\eqref{eq:Boltz.graph.migr} of homogeneous Boltzmann-type kinetic equations is richer than system~\eqref{eq:rhoi} for the sole mass redistribution, as it can provide information on the detailed distribution of the microscopic trait $v$ of the agents on the graph. Its analytical study is also more challenging, as it requires more advanced tools than those underlying the results discussed in Section~\ref{sect:mass_redistribution}.

One of such tools is the \textit{Fourier metric} for probability measures, which, as first noticed by Bobylev~\cite{Bobylev1975}, is particularly effective in producing \textit{a priori} estimates for Boltzmann-type equations with linear interactions. To introduce the topic, let $\mu$ be a probability measure on $\R$ equipped with e.g., the Borel $\sigma$-algebra. For $\xi\in\R$, the bounded and continuous function $\hat{\mu}:\R\to\R$ defined as
$$ \hat{\mu}(\xi):=\int_\R e^{-i\xi v}\,d\mu(v) $$
is the Fourier transform of the measure $\mu$. If $\mu,\,\nu$ are two probability measures on $\R$ such that $\int_\R\abs{v}^s\,d\mu(v)<+\infty$ and likewise $\int_\R\abs{v}^s\,d\nu(v)<+\infty$ for some real $s>0$ then
\begin{equation}
	d_s(\mu,\nu):=\sup_{\xi\in\R\setminus\{0\}}\frac{\abs{\hat{\nu}(\xi)-\hat{\mu}(\xi)}}{\abs{\xi}^s}
	\label{eq:ds}
\end{equation}
defines a metric, called the \textit{$s$-Fourier metric}, which measures the distance between $\mu$ and $\nu$. Notice that the well-posedness of $d_s(\mu,\nu)$ relies heavily on the fact $\mu,\,\nu$ are probability measures, which guarantees $\hat{\mu}(0)=\hat{\nu}(0)=1$. This, together with additional properties on the statistical moments of $\mu,\,\nu$,  is at the basis of the finiteness of the right-hand side of~\eqref{eq:ds} when $\xi$ is close to $0$, cf.~\cite{Carrillo2007,Loy2025} for details.

Applying the Fourier transform to both sides of~\eqref{eq:Boltz.graph.migr} simplifies greatly the Boltzmann-type equation. In fact, taking advantage of the assumed linearity of the interaction rules, it results
$$ \int_\R e^{-i\xi v}Q(f_i,f_i)(v,t)\,dv=\hat{f}_i(p\xi,t)\hat{f}_i(q\xi,t)-\hat{f}_i(\xi,t), $$
cf.~\cite{Loy2025}, thus~\eqref{eq:Boltz.graph.migr} becomes
\begin{equation}
	\frac{\partial\hat{f}_i}{\partial t}(\xi,t)=\lambda\left(\hat{f}_i(p\xi,t)\hat{f}_i(q\xi,t)-\hat{f}_i(\xi,t)\right)+\chi\left(\sum_{j=1}^{N}a_{ij}\hat{f}_j(\xi,t)-\hat{f}_i(\xi,t)\right),
		\qquad i=1,\,\dots,\,N.
	\label{eq:Boltz.Fourier_transf}
\end{equation}
Ideally, this transformed equation is the starting point for estimates in the Fourier metric. Nevertheless, unlike the standard kinetic setting, here the $f_i$'s are not probability measures, in fact $\int_\R f_i(v,t)\,dv=\rho_i(t)\leq 1$. Hence, definition~\eqref{eq:ds} cannot be applied straightforwardly.

To bypass this difficulty of the theory, it is useful to introduce the representation
$$ f_i(v,t)=\rho_i(t)F_i(v,t), \qquad i=1,\,\dots,\,N, $$
where $F_i:\R\times (0,\,+\infty)\to\R_+$ is the so-called ($i$-th) \textit{normalised kinetic distribution function}. If we confine ourselves to the case in which the initial mass distribution is strictly positive in every vertex of $\cG_N$, i.e. $\rho_{0,i}>0$ for all $i=1,\,\dots,\,N$, then Theorems~\ref{theo:rho},~\ref{theo:rho.asympt}
imply that the $F_i$'s are probability distributions, as $\int_\R F_i(v,t)\,dv=1$ for all $t>0$ and all $i=1,\,\dots,\,N$. Consequently, the quantity
\begin{equation}
	D_s(\bbf(t),\bbg(t)):=\sum_{i=1}^{N}\rho_i(t)d_s(F_i(t),G_i(t))
	\label{eq:Ds}
\end{equation}
turns out to be a metric for every pair of vector-valued kinetic distributions
\begin{align*}
	\bbf(v,t) &:= (f_1(v,t),\,\dots,\,f_N(v,t))=(\rho_1(t)F_1(v,t),\,\dots,\,\rho_N(t)F_N(v,t)), \\
	\bbg(v,t) &:= (g_1(v,t),\,\dots,\,g_N(v,t))=(\rho_1(t)G_1(v,t),\,\dots,\,\rho_N(t)G_N(v,t))
\end{align*}
having the same mass $\brho(t)=(\rho_1(t),\,\dots,\,\rho_N(t))$ on the graph. Notice that $D_s$ is well defined because it is based on the Fourier distances between the $F_i$'s and $G_i$'s, which are probability distributions by construction, and because the weighting coefficients in the sum~\eqref{eq:Ds} are strictly positive in our context. Notice also that it is quite easy to ensure that two kinetic distributions $\bbf$, $\bbg$ have the same mass in the vertices of the graph: owing to~\eqref{eq:rhoi}, it is sufficient that initially they have the same mass distribution to get that their mass distributions coincide at all successive times.

In this setting, under technical assumptions reported in detail in~\cite{Loy2025}, which involve, among other things, the rate $\lambda$ in~\eqref{eq:Boltz.graph.migr} and the coefficients $p,\,q$ of the interaction rules~\eqref{eq:int_rules}, one proves that there exists a constant $\gamma>0$ such that
\begin{equation}
	D_2(\bbf(t),\bbg(t))\leq D_2(\bbf_0,\bbg_0)e^{-\gamma t}, \qquad \forall\,t>0,
	\label{eq:D2}
\end{equation}
where $\bbf_0(v):=\bbf(v,0)$ and $\bbg_0(v):=\bbg(v,0)$ are the initial kinetic distributions. The choice of $s=2$ in~\eqref{eq:D2} as the index of the Fourier metric is linked to the assumptions mentioned above on the coefficients $p,\,q$ of the interaction rules.

It is worth stressing that~\eqref{eq:D2} is not just a continuous dependence estimate for the solutions of~\eqref{eq:Boltz.graph.migr}. It says, more specifically, that
$$ \lim_{t\to +\infty}D_2(\bbf(t),\bbg(t))=0, $$
meaning that solutions to~\eqref{eq:Boltz.graph.migr} tend to exhibit the same time-asymptotic trend regardless of the initial data. In turn, this implies the \textit{a priori} uniqueness and global asymptotic stability of the equilibrium distribution of the trait $v$ emerging, in the long run, from the joint effect of intra-vertex agent interactions and inter-vertex agent migrations. Remarkably, in this context such an equilibrium distribution is the equivalent of the Maxwellian distribution for the homogeneous Boltzmann equation of gas dynamics.

To complete the theoretical big picture, we mention that by means of other more standard metrics, such as the $N$-dimensional Lebesgue metrics induced by the norms
$$ \norm{\bbf(t)}{(L^r(\R))^N}:={\left(\sum_{i=1}^{N}\norm{f_i(t)}{L^r(\R)}^r\right)}^{1/r}
	={\left(\sum_{i=1}^{N}\int_\R f_i^r(v,t)\,dv\right)}^{1/r}, \qquad r=1,\,2, $$
one proves an \textit{a priori} continuous dependence and uniqueness estimate for the solutions of~\eqref{eq:Boltz.graph.migr} in the form
$$ \norm{\bbg(t)-\bbf(t)}{(L^2(\R))^N}\leq{\left(\norm{\bbg_0-\bbf_0}{(L^2(\R))^N}^2+\psi(t)\norm{\bbg_0-\bbf_0}{(L^1(\R))^N}\right)}^{1/2}e^{Ct}, \qquad \forall\,t>0, $$
which is technically obtained from the Fourier-transformed system~\eqref{eq:Boltz.Fourier_transf} invoking Parseval's identity and does not require any of the technical assumptions on $\lambda$, $p$, $q$, and the initial mass distribution mentioned above. Here, $\psi$ is a suitable non-negative and non-decreasing function with $\psi(0)=0$ and, more importantly, the real constant $C$ in the exponential function is positive. Therefore, unlike~\eqref{eq:D2}, such an estimate, although fundamental for the well-posedness of the problem, does not provide insights into the time-asymptotic trend of the solutions.

\section{Networked interactions}
\label{sect:networked_int}
\begin{figure}[!t]
\centering
\includegraphics[width=.45\textwidth]{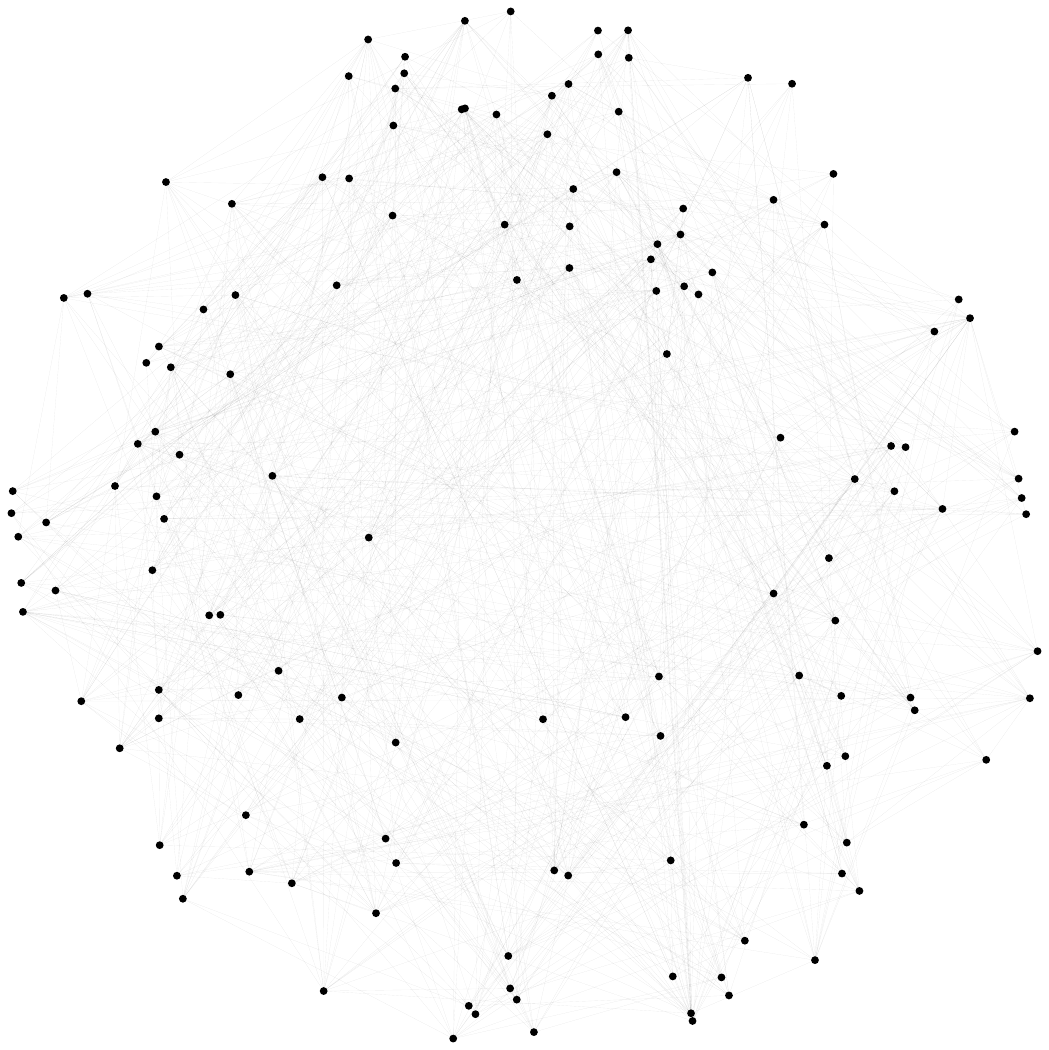}
\caption{A dense graph of interactions in a multi-agent system}
\label{fig:dense_graph}
\end{figure}

We consider now the case of \textit{networked interactions}, that depict a situation in which the agents of the multi-agent system are the vertices of $\cG_N$ and interactions among them may or may not take place depending on the edge structure of $\cG_N$. In this case, therefore, there is formally a single multi-agent system, whose individuals are not subject, in general, to an ``all-to-all'' interaction pattern. In fact, only agents sharing a connection can possibly interact, whereby they exchange mutually their respective traits. See Figure~\ref{fig:dense_graph}.

Networked interactions are ubiquitous in sociophysics and biophysics. Probably the most prominent example is the opinion formation on social networks, see e.g.,~\cite{Albi2016,Albi2017,Burger2025,Duering2024,Fraia2026,Loy2022,Nurisso2024,Toscani2018}. Moreover, we mention that models of the brain network based on networked interactions among neurons have been proposed in the literature to describe the progression of Alzheimer's disease~\cite{Bertsch2017} and the activation/deactivation of neurons characterising the electrical brain activity~\cite{Bisi2026,Conte2024}. Recently, kinetic models on networks have also been employed to investigate the city size distribution subject to immigration and emigration phenomena~\cite{Taralli2026}.

On a heuristic basis, considering a system composed by a large number of agents -- viz. a graph $\cG_N$ with large $N$ --, one describes the microscopic state of a representative agent by means of a pair $(v,\,c)\in\R\times\R_+$, where $v$ denotes the physical trait of the agent, such as e.g., their opinion, and $c$ is the \textit{number of connections} with other agents. Then one introduces the kinetic distribution function $f=f(v,c,t):\R\times\R_+\times (0,\,+\infty)\to\R_+$ such that $f(v,c,t)\,dv\,dc$ is the probability that at time $t$ the microscopic state of the representative agent is in the infinitesimal volume $[v-\frac{dv}{2},\,v+\frac{dv}{2}]\times [c-\frac{dc}{2},\,c+\frac{dc}{2}]$ of the state space $\R\times\R_+$ centred at $(v,\,c)$. In this context, conservation of the total mass of the agents holds true, which motivates the normalisation condition
$$ \int_{\R_+}\int_\R f(v,c,t)\,dv\,dc=1, \qquad \forall\,t\geq 0. $$
A common assumption, which we will make throughout this section, is that the network is \textit{non-coevolving}, meaning that the edges of $\cG_N$ do not vary in time. This implies that the marginal distribution, say $g$, of the connections:
$$ g(c,t):=\int_\R f(v,c,t)\,dv, $$
is constant in $t$ and will be therefore denoted simply by $g(c)$. The interested reader may refer e.g., to~\cite{Burger2022,Fraia2026a} for the case of \textit{coevolving networks}, i.e. networks with time-varying connections.

Considering prototypical interaction rules of the form~\eqref{eq:inv_int}, a general Boltzmann-type description of such a multi-agent system is provided by the equation
\begin{equation}
	\frac{\partial f}{\partial t}(v,c,t)=\int_{\R_+}\int_\R B(c,c_\ast)\left(\frac{1}{\abs{p^2-q^2}}f(\pr{v},c,t)f(\pr{v}_\ast,c_\ast,t)-f(v,c,t)f(v_\ast,c_\ast,t)\right)dv_\ast\,dc_\ast,
	\label{eq:Boltz.c}
\end{equation}
where the interaction kernel $B$, giving the rate of interaction, is assumed to depend on the number of connections of the interacting agents but not on their traits. For example, in~\cite{Loy2022} it is suggested that $B$ may be proportional to the product $cc_\ast$, implying that the more the connections the more frequent the interactions.

Integrating~\eqref{eq:Boltz.c} with respect to $v$ produces
$$ \frac{\partial g}{\partial t}(c,t)=\int_{\R_+}B(c,c_\ast)\bigl(g(c,t)g(c_\ast,t)-g(c,t)g(c_\ast,t)\bigr)\,dc_\ast=0, $$
thereby confirming that the connection distribution is conserved in time. In practice, it can be fixed by prescribing the initial condition $f_0(v,c):=f(v,c,0)$ in such a way that $\int_\R f_0(v,c)\,dv=g(c)$. As for the choice of $g$, the literature offers several studies on the statistical distribution of graph connections, see e.g.,~\cite{Albert2002,Barabasi1999,Barabasi1999a}. In particular, it is interesting to notice that \textit{power-law-type} distributions $g$, i.e. such that $g(c)\sim c^{-\alpha}$ for some $\alpha>1$ when $c\to +\infty$, can be used to model social networks featuring the so-called \textit{influencers}, i.e. agents able to affect significantly the opinion of other agents thanks to their large number of connections. An example is
$$ g(c)=\frac{e^{-1/c}}{\Gamma(\alpha-1)c^\alpha}, \qquad c>0,\quad\alpha>1, $$
which, for $c$ large, is a power law of degree $\alpha$. Here, $\Gamma$ denotes the Euler gamma function. Conversely, distributions $g$ such that $g(c)=o(c^{-\alpha})$ for every $\alpha>1$ when $c\to +\infty$ can be used to model social networks substantially free from influencers. An example of this type is
$$ g(c)=\mu e^{-\mu c}, \qquad c\geq 0, $$
namely an exponential distribution with parameter $\mu>0$.

\subsection{Formal derivation of a Boltzmann-type equation on a graph}
Equation~\eqref{eq:Boltz.c} is built by mimicking heuristically the classical homogeneous Boltzmann-type equation~\eqref{eq:homog_Boltz-type}. In particular, the graph $\cG_N$ is directly blurred into a continuous statistical description of its connections, with no explicit link to the genuinely discrete structure of vertices and edges conveyed by the adjacency matrix. In this section, we want to restore such a link by discussing how a statistical description of a graph can be formally embedded into a homogeneous Boltzmann-type equation in the limit of an infinite number of vertices.

As stated in the introductory section, we recall that we consider for simplicity undirected graphs $\cG_N$ with symmetric adjacency matrices $\bbA_N$. Hence, if an edge exists between two vertices then the corresponding agents can interact and exchange their traits reciprocally via the symmetric interaction rules~\eqref{eq:int_rules},~\eqref{eq:inv_int}.

Let $f_i=f_i(v,t):\R\times (0,\,+\infty)\to\R_+$ be the distribution function of the trait $v$ of vertex $i$ of $\cG_N$ at time $t$. Taking into account the connections of vertex $i$ with the other vertices of $\cG_N$ as expressed by $\bbA_N$, the distribution function $f_i$ satisfies~\cite{Nurisso2024}:
\begin{equation}
	\frac{\partial f_i}{\partial t}(v,t)=\frac{1}{N}\sum_{j=1}^{N}a_{ij}\int_\R\left(\frac{1}{\abs{p^2-q^2}}f_i(\pr{v},t)f_j(\pr{v}_\ast,t)-f_i(v,t)f_j(v_\ast,t)\right)dv_\ast,
	\label{eq:fi}
\end{equation}
see also~\cite{Burger2021}, where a similar kinetic equation is derived from a BBGKY-type hierarchy. In weak form, for every observable $\varphi=\varphi(v):\R\to\R$ (test function), this writes
\begin{equation}
	\frac{d}{dt}\int_\R\varphi(v)f_i(v,t)\,dv=\frac{1}{N}\sum_{j=1}^{N}a_{ij}\int_\R\int_\R\bigl(\varphi(v')-\varphi(v)\bigr)f_i(v,t)f_j(v_\ast,t)\,dv\,dv_\ast,
	\label{eq:fi.weak}
\end{equation}
cf.~\eqref{eq:homog_Boltz-type.weak}, and, letting $\bbf(v,t):=(f_1(v,t),\,\dots,\,f_N(v,t))$, further
$$ \frac{d}{dt}\int_\R\varphi(v)\bbf(v,t)\,dv=\frac{1}{N}\int_\R\int_\R\bigl(\varphi(v')-\varphi(v)\bigr)\bbf(v,t)\odot\bbA_N\bbf(v_\ast,t)\,dv\,dv_\ast, $$
where $\odot$ denotes the Hadamard product, i.e. the entrywise product of vectors.

Let now
$$ c_i:=\deg{(i)}\in\{0,\,\dots,\,N\} $$
be the \textit{degree} of vertex $i$, i.e. the number of incoming/outgoing edges of $i$ (for every $i$, these two numbers are equal because the graph is undirected) and set $\bbc:=(c_1,\,\dots,\,c_N)^T$ (column vector -- the superscript $T$ denoting transposition). Aiming to retain only the information about the number of connections of every vertex, we introduce the following approximation of the adjacency matrix $\bbA_N$:
\begin{equation}
	\bbA_N\approx\frac{\bbc\bbc^T}{d_N},
		\qquad d_N:=\sum_{i=1}^{N}\deg{(i)}=\sum_{i=1}^{N}c_i.
	\label{eq:AN}
\end{equation}
The right-hand side of the approximation $\approx$ is a rank-one $N\times N$ matrix with the same vertex degrees as $\bbA_N$. Such an approximate adjacency matrix corresponds, in practice, to replacing $\cG_N$ with a \textit{random graph} with the same degree sequence as $\cG_N$, cf. the \textit{configuration model}~\cite{Bollobas1980,Molloy1995}. In fact, we observe that, on the one hand, the approximation~\eqref{eq:AN} of $\bbA_N$ loses the detailed structure of the connections of $\cG_N$. On the other hand, it amounts to the following approximation of the generic entry $a_{ij}$ of $\bbA_N$:
$$ a_{ij}\approx\frac{c_ic_j}{d_N}, \qquad i,\,j=1,\,\dots,\,N, $$
whence the exact degree of vertex $i$ of $\cG_N$, i.e. $c_i=\sum_{j=1}^{N}a_{ij}$, coincides with the approximate degree $\sum_{j=1}^{N}c_ic_j/d_N=c_i$. Interestingly, random graphs with a prescribed degree sequence have proved particularly suitable for modelling social and Internet networks.

In parallel, we introduce the distribution function
$$ f=f(v,c,t):\R\times\{0,\,\dots,\,N\}\times (0,\,+\infty)\to\R_+ $$
such that $f(v,c,t)\,dv$ is the probability that at time $t$ a vertex/agent has $c$ connections and a trait $v$ comprised in $[v-\frac{dv}{2},\,v+\frac{dv}{2}]$:
$$ f(v,c,t):=\frac{1}{N}\sum_{i\,:\,\deg{(i)}=c}f_i(v,t). $$
From this definition it follows, in particular,
$$ \sum_{c=0}^{N}f(v,c,t)=\frac{1}{N}\sum_{i=1}^{N}f_i(v,t), \qquad
	\sum_{c=0}^{N}cf(v,c,t)=\frac{1}{N}\sum_{i=1}^{N}c_if_i(v,t). $$
Therefore, summing both sides of~\eqref{eq:fi.weak} over $i$ while using~\eqref{eq:AN} therein, we get
\begin{equation}
	\frac{d}{dt}\sum_{c=0}^{N}\int_\R\varphi(v)f(v,c,t)\,dv=\frac{1}{d_N}\sum_{c=0}^{N}\sum_{c_\ast=0}^{N}
		\int_\R\int_\R cc_\ast\bigl(\varphi(v')-\varphi(v)\bigr)f(v,c,t)f(v_\ast,c_\ast,t)\,dv\,dv_\ast.
	\label{eq:Boltz.graph.f.c}
\end{equation}

At this point, to obtain a limiting mathematical structure as $N\to\infty$ it is convenient to normalise the degree of the vertices as
$$ \tilde{c}:=\frac{c}{N}, $$
so that the normalised degree $\tilde{c}$ takes values between $0$ and $1$ in the set $\tilde{\cC}_N:=\{\frac{k}{N}\,:\,k=0,\,\dots,\,N\}$. At the same time, we normalise also the distribution function $f$ by letting
$$ \tilde{f}(v,\tilde{c},t):=Nf(v,N\tilde{c},t). $$
For $v$, $t$ fixed, it is useful to understand $\tilde{f}(v,\cdot,t)$ as a piecewise constant version of $f(v,\cdot,t)$ in the interval $[0,\,1]$, after discretising the latter by means of the grid $\tilde{\cC}_N$ with grid step $\Delta{\tilde{c}}:=\frac{1}{N}$. In contrast, $f(v,\cdot,t)$ can be viewed as an atomic measure on the discrete set $\{0,\,\dots,\,N\}$.

Switching to $\tilde{f}$ in~\eqref{eq:Boltz.graph.f.c} we discover
\begin{multline*}
	\frac{d}{dt}\sum_{\tilde{c}\in\tilde{\cC}_N}\int_\R\varphi(v)\tilde{f}(v,\tilde{c},t)\,dv\,\Delta{\tilde{c}} \\
	=\frac{N^2}{d_N}\sum_{\tilde{c}\in\tilde{\cC}_N}\sum_{\tilde{c}_\ast\in\tilde{\cC}_N}\int_\R\int_\R\tilde{c}\tilde{c}_\ast
		\bigl(\varphi(v')-\varphi(v)\bigr)\tilde{f}(v,\tilde{c},t)\tilde{f}(v_\ast,\tilde{c}_\ast,t)\,dv\,dv_\ast\,\Delta{\tilde{c}}\,\Delta{\tilde{c}_\ast}.
\end{multline*}
It is clear now that, when $N\to\infty$, the variables $\tilde{c}$, $\tilde{c}_\ast$ span continuously the interval $[0,\,1]$ whereas expressions such as $\sum_{\tilde{c}\in\tilde{\cC}_N}(\dots)\,\Delta{\tilde{c}}$ converge formally to $\int_0^1(\dots)\,d\tilde{c}$, cf. the construction of the Riemann integral. Moreover, the ratio $d_N/N^2$ converges to the \textit{mean normalised degree} of the graph:
$$ \frac{d_N}{N^2}\xrightarrow{N\to\infty}\tilde{d}:=\int_0^1\int_\R\tilde{c}\tilde{f}(v,\tilde{c},t)\,dv\,d\tilde{c}, $$
as proved in~\cite{Nurisso2024}.

In conclusion, in the limit $N\to\infty$ the following equation emerges:
\begin{multline*}
	\frac{d}{dt}\int_0^1\int_\R\varphi(v)\tilde{f}(v,\tilde{c},t)\,dv\,d\tilde{c} \\
	=\int_0^1\int_0^1\int_\R\int_\R\frac{\tilde{c}\tilde{c}_\ast}{\tilde{d}}\bigl(\varphi(v')-\varphi(v)\bigr)
		\tilde{f}(v,\tilde{c},t)\tilde{f}(v_\ast,\tilde{c}_\ast,t)\,dv\,dv_\ast\,d\tilde{c}\,d\tilde{c}_\ast,
\end{multline*}
which, owing to the arbitrariness of $\varphi$, is a weak form of the homogeneous Boltzmann-type equation
\begin{equation}
	\frac{\partial\tilde{f}}{\partial t}(v,\tilde{c},t)=\int_0^1\int_\R\frac{\tilde{c}\tilde{c}_\ast}{\tilde{d}}
		\left(\frac{1}{\abs{p^2-q^2}}\tilde{f}(\pr{v},\tilde{c},t)\tilde{f}(\pr{v}_\ast,\tilde{c}_\ast,t)
			-\tilde{f}(v,\tilde{c},t)\tilde{f}(v_\ast,\tilde{c}_\ast,t)\right)dv_\ast\,d\tilde{c}_\ast
	\label{eq:Boltz.f_tilde}
\end{equation}
for the interaction rules~\eqref{eq:inv_int}. This equation resembles closely~\eqref{eq:Boltz.c} with the specific ``non-Maxwellian'', i.e. non-unitary, interaction kernel
$$ B(\tilde{c},\tilde{c}_\ast)=\frac{\tilde{c}\tilde{c}_\ast}{\tilde{d}}. $$
The most remarkable difference is that here the normalised numbers of connections are used, which range over $[0,\,1]$ rather than over $\R_+$. From the previous discussion, it is clear that such a normalisation has been essential to identify a limiting equation for $N\to\infty$.

In summary,~\eqref{eq:Boltz.f_tilde} describes networked interactions on large \textit{random} graphs with a prescribed non-coevolving statistical distribution $\tilde{g}(\tilde{c})=\int_\R\tilde{f}(v,\tilde{c},t)\,dv$ of the (normalised) connections. 

\subsection{Embedding graphons in Boltzmann-type equations}
\begin{figure}[!t]
\centering
\includegraphics[width=.5\textwidth]{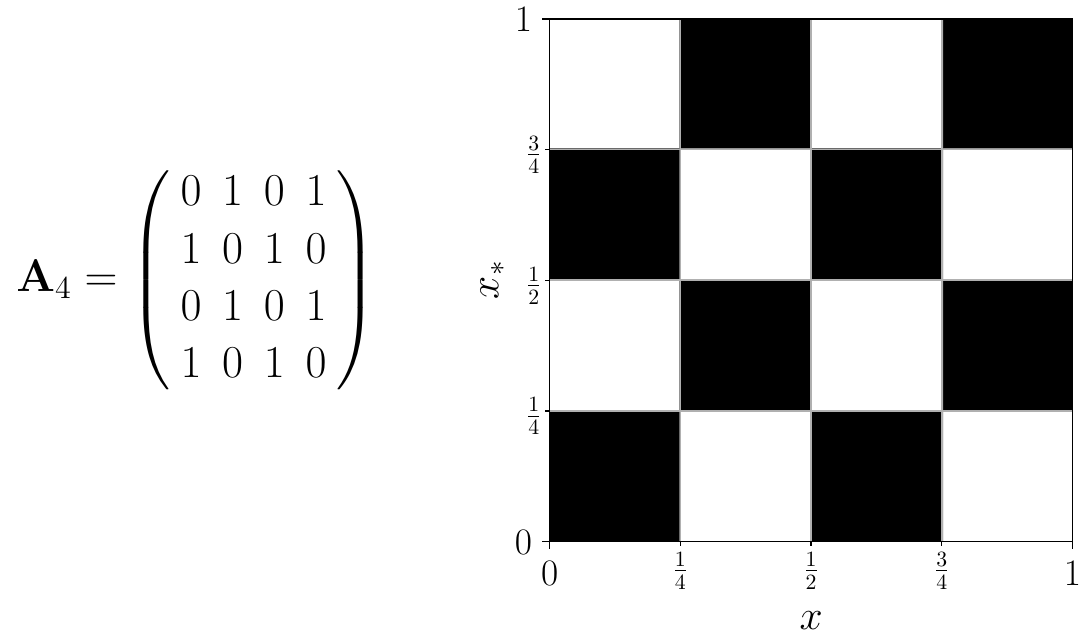}
\caption{A $4\times 4$ adjacency matrix and the contours of the corresponding piecewise constant function $W_4$ on the unit square $[0,\,1]^2$ (white is $0$, black is $1$)}
\label{fig:WN}
\end{figure}

By revisiting the approach that led to~\eqref{eq:fi.weak} and~\eqref{eq:Boltz.f_tilde}, we can obtain a different homogeneous Boltzmann-type equation, which accounts for networked interactions on large but not necessarily random graphs. In particular, such an equation keeps track of the structure of the graph connections.

To this purpose, the main concept to be borrowed from graph theory is that of \textit{graphon}, a name inspired by ``\textit{graph} functi\textit{on}'' which became standard in the mid-2000s after the works by Lov\'{a}sz and collaborators, see e.g.,~\cite{Lovasz2012,Lovasz2006}.

In essence, the idea behind a graphon works as follows. Given a finite graph $\cG_N$ with $N$ vertices and $N\times N$ adjacency matrix $\bbA_N$, one first introduces a partition of the interval $[0,\,1]\subset\R$ made by the subintervals
$$ I_1:=\left[0,\,\frac{1}{N}\right], \qquad I_i:=\left(\frac{i-1}{N},\,\frac{i}{N}\right], \quad i=2,\,\dots,\,N, $$
which are such that $\cup_{i=1}^{N}I_i=[0,\,1]$ and $I_i\cap I_j=\emptyset$ for all $i\neq j$, and then defines a two-variable scalar function $W_N:[0,\,1]^2\to [0,\,1]$ as
$$ W_N(x,x_\ast):=\sum_{i=1}^{N}\sum_{j=1}^{N}a_{ij}\chi_{I_i}(x)\chi_{I_j}(x_\ast), $$
    where $\chi_{(\cdot)}$ denotes the characteristic function of the set specified by the subscript. In practice, $W_N$ is a piecewise constant function reproducing $\bbA_N$ on a pixelation of the unit square $[0,\,1]^2\subset\R^2$, in the sense that
$$ W_N(x,x_\ast)=a_{ij} \quad \text{for} \quad (x,\,x_\ast)\in I_i\times I_j, $$
$I_i\times I_j$ being a sub-square, viz. a pixel, in $[0,\,1]^2$. See Figure~\ref{fig:WN} for an example.

As the graph $\cG_N$ gets larger and larger, in particular for $N\to\infty$, one may expect the sequence of functions $\{W_N\}_{N\in\N}$ to converge, under a suitable concept of convergence to be defined, to some limit function $W:[0,\,1]^2\to [0,\,1]$ expressing the connectivity of $\cG_N$ in the limit of an infinite number of vertices. Such a function $W$ is the so-called graphon.

While the entry $a_{ij}$ of the adjacency matrix $\bbA_N$ describes whether the vertices $i$ and $j$ of $\cG_N$ are ($a_{ij}=1$) or are not ($a_{ij}=0$) connected by an edge, the value $W(x,x_\ast)\in [0,\,1]$ is typically interpreted as the probability of an edge between the two points $x,\,x_\ast\in [0,\,1]$, where, owing to the construction set forth above, the latter are understood as a continuous counterpart of the (normalised) vertices of $\cG_N$.

In the context of graphons, the most popular concept of convergence is that induced by the so-called \textit{cut norm}, which is denoted by $\norm{\cdot}{\square}$ and is defined as
$$ \norm{W}{\square}:=\sup_{S,\,T\subseteq [0,\,1]}{\abs*{\iint_{S\times T}W(x,x_\ast)\,dx\,dx_\ast}},
	\qquad W\in L^\infty([0,\,1]^2). $$
When applied to the function $W_N$ introduced above, the cut norm yields
$$ \norm{W_N}{\square}=\sup_{S,\,T\subseteq [0,\,1]}\abs*{\sum_{i=1}^{N}\sum_{j=1}^{N}a_{ij}\abs{S\cap I_i}\cdot\abs{T\cap I_j}}, $$
where $\abs{S\cap I_i}$ is the Lebesgue measure of the set $S\cap I_i$ and similarly for $\abs{T\cap I_j}$. From here, we see that, informally speaking, the cut norm scans the sub-regions of a graph, counting the number of edges between pairs of vertices in those sub-regions, to detect the maximum \textit{density of connections} of the graph.

As usual, we say that the sequence $\{W_N\}_{N\in\N}$ converges to $W$ in the cut norm when $N\to\infty$ if $\lim_{N\to\infty}\norm{W_N-W}{\square}=0$.

Given a multi-agent system with networked interactions taking place on $\cG_N$, we denote by $f_N=f_N(x,v,t)$ the probability density function of the pair $(x,\,v)\in [0,\,1]\times\R$ at time $t>0$. Thus, $f_N(x,v,t)\,dx\,dv$ is the probability that at time $t$ an agent/vertex is in the sub-region $[x-\frac{dx}{2},\,x+\frac{dx}{2}]$ of the graph centred at $x$ with a trait in $[v-\frac{dv}{2},\,v+\frac{dv}{2}]$. On the whole,
$$ \int_0^1\int_\R f_N(x,v,t)\,dv\,dx=1, \qquad \forall\,t>0, \quad \forall\,N\in\N. $$
Following~\cite{Taricco2026}, thus using in particular the function $W_N$ to express the connections among the agents/vertices, we write an analogue of~\eqref{eq:fi} in the form
\begin{multline*}
	\frac{\partial f_N}{\partial t}(x,v,t) \\
	=\int_0^1\int_\R W_N(x,x_\ast)
		\left(\frac{1}{\abs{p^2-q^2}}f_N(x,\pr{v},t)f_N(x_\ast,\pr{v}_\ast,t)-f_N(x,v,t)f_N(x_\ast,v_\ast,t)\right)dv_\ast\,dx_\ast
\end{multline*}
for the interaction rules~\eqref{eq:inv_int}, where the discrete indices $i,\,j$ and the entries $a_{ij}$ of $\bbA_N$ have been replaced by the continuous variables $x,\,x_\ast$ and the function $W_N$, respectively. Similarly to~\eqref{eq:fi}, this equation constitutes a kinetic description of networked interactions on a ``physical'' graph, namely a finite one with a pointwise structure of edges and connections.

In the limit $N\to\infty$, if the sequence $\{W_N\}_{N\in\N}$ converges to a graphon $W$ it is reasonable to expect that the distribution function $f_N$ converges to the solution $f$ of the following limit homogeneous Boltzmann-type equation:
\begin{multline}
	\frac{\partial f}{\partial t}(x,v,t) \\
	=\int_0^1\int_\R W(x,x_\ast)
		\left(\frac{1}{\abs{p^2-q^2}}f(x,\pr{v},t)f(x_\ast,\pr{v}_\ast,t)-f(x,v,t)f(x_\ast,v_\ast,t)\right)dv_\ast\,dx_\ast.
	\label{eq:Boltz.graphon}
\end{multline}
Such an intuition is made precise by the following \textit{a priori} estimate proved in~\cite{Taricco2026}:
\begin{equation}
	\sup_{t\in [0,\,T]}\cW_1(f_N(t),f(t))\lesssim\cW_1(f_N(0),f(0))+\norm{W_N-W}{\square}^{1/2}+\sqrt{N}\norm{W_N-W}{\square},
		\quad\forall\,T>0,
	\label{eq:estimate.W1}
\end{equation}
where $\cW_1$ denotes the \textit{1-Wasserstein metric} for probability measures:
$$ \cW_1(f_N(t),f(t)):=\sup_{\Phi\in\operatorname{Lip}_1([0,\,1]\times\R)}\iint_{[0,\,1]\times\R}\Phi(x,v)\bigl(f_N(x,v,t)-f(x,v,t)\bigr)\,dx\,dv, $$
being $\operatorname{Lip}_1([0,\,1]\times\R)$ the set of Lipschitz-continuous functions on $[0,\,1]\times\R$ with at most unitary Lipschitz constant.

Estimate~\eqref{eq:estimate.W1} shows that if the $W_N$'s converge to a graphon $W$ \textit{sufficiently fast} in the cut norm, in particular in such a way that $\norm{W_N-W}{\square}=o(N^{-1/2})$ when $N\to\infty$, and if, in addition to this, the initial conditions $f_N(x,v,0)$ converge to some $f(x,v,0)$ then the $f_N$'s converge, uniformly in time on every compact interval $[0,\,T]$, to the solution of~\eqref{eq:Boltz.graphon} issuing from $f(x,v,0)$.

This establishes that~\eqref{eq:Boltz.graphon} is the limit kinetic model valid in the abstraction of an infinite graph described by a graphon. As discussed above, it also clarifies the structural assumptions required for~\eqref{eq:Boltz.graphon} to be regarded as a \textit{universal} model for families of growing finite graphs.

Notice that~\eqref{eq:Boltz.graphon} is a homogeneous Boltzmann-type equation on the state space $[0,\,1]\times\R$ with the graphon $W$ playing the role of the interaction kernel $B$. With this interpretation, a similar kinetic equation has been proposed heuristically in~\cite{Duering2024} for a problem of opinion formation on social networks.

\bibliographystyle{plain}
\bibliography{biblio}
\end{document}